\documentclass[%
 reprint,
 amsmath,amssymb,
 aps,
]{revtex4-2}

\usepackage{graphicx}
\usepackage{dcolumn}
\usepackage{bm}
\usepackage{float}
\usepackage{natbib}

\begin{document}

\preprint{APS/123-QED}

\title{Quantum Error Detection Without Using Ancilla Qubits}

\author{Nicolas J. Guerrero}
\author{David E. Weeks}
 \email{david.weeks@afit.edu}
\affiliation{%
 Air Force Institute of Technology
}%

\date{\today}

\begin{abstract}
In this paper, we describe and experimentally demonstrate an error detection scheme that does not employ ancilla qubits or mid-circuit measurements. This is achieved by expanding the Hilbert space where a single logical qubit is encoded using several physical qubits. For example, one possible two qubit encoding identifies $|0\rangle_L=|01\rangle$ and $|1\rangle_L=|10\rangle$. If during the final measurement a $|11\rangle$ or $|00\rangle$ is observed an error is declared and the run is not included in subsequent analysis. We provide codewords for a simple bit-flip encoding, a way to encode the states, a way to implement logical $U_3$ and logical $C_x$ gates, and a description of which errors can be detected. We then run Greenberger–Horne–Zeilinger circuits on the transmon based IBM quantum computers, with an input space of $N\in\{2,3,4,5\}$ logical qubits and $Q\in\{1,2,3,4,5\}$ physical qubits per logical qubit. The results are compared relative to $Q=1$ with and without error detection and we find a significant improvement for $Q\in\{2,3,4\}$.

\end{abstract}

\maketitle


\section{\label{sec:level1} Introduction}

In an ideal world, physical qubits of quantum computers would be isolated from outside influence to permit useful and reliable quantum computation. Of course, it is well understood that such perfect isolation is likely an impossibility, and that there are many things interacting with quantum computers which prevent perfect qubits. These include, but are not limited to quantum vacuum fluctuations \cite{QuantumVacuum3,QuantumVacuum1}, cosmic radiation \cite{Cosmic1,Cosmic2,Cosmic3}, material defects \cite{defect1,defect2,defect3}, and many others.

The most promising idea to mitigate these errors is currently quantum error correction (QEC) and quantum error detection (QED). The first error correction code employed nine physical qubits per logical qubit and six ancilla qubits for syndrome extraction \cite{Correct9}. More efficient codes were soon produced such as codes on seven qubits \cite{Correct7} and five qubits \cite{Correct5}, as well as other error correction schemes such as surface codes \cite{Surface1} and continuous QEC \cite{Con1,Con2}. These codes have been studied extensively through simulations \cite{Sim7,Sim579,Sim5} as well as more recently through experimentation \cite{Exp1,Exp2,Exp3,Exp4,Exp5}.

In a similar manner, QED was first described using four physical qubits per logical qubit and two ancilla qubits for syndrome extraction \cite{Detect1}. Other types of detection process exist such as QED in surface codes \cite{Detect3} and hardware based sensors \cite{Detect2}. QED has been experimentally verified for small numbers of logical qubits \cite{Detect3,Detect4} and continues to be a useful companion to QEC. In this paper, we will introduce and experimentally verify a new type of quantum error detection which differs significantly from both QEC and QED. The experimental verification is conducted on the transmon based IBM quantum computers \cite{ibmquantum}.

\subsection{No ancilla error detection}

No ancilla error detection (NAED) is a comparable process to both QEC and QED. In all three, $Q$ physical qubits represent a single logical qubit with states $|0\rangle_L$ and $|1\rangle_L$. NAED differs from QEC and the standard QED in the rather obvious way that it does not use ancilla qubits. This has two benefits over previous schemes: the resources required are generally less than QEC and QED and mid-circuit measurements are not a requirement for NAED. The only measurements are made at the end of the circuit, thereby bypassing any noise that might otherwise be incurred by measuring ancilla qubits.

An implementation of NAED begins by considering a single logical qubit $|\psi\rangle_L=\alpha|0\rangle_L+\beta|1\rangle_L$ over $Q$ physical qubits. Define the set $H^{+}=\{\alpha|0\rangle_L+\beta|1\rangle_L:|\alpha|^2+|\beta|^2=1\}$ and define the set $H^{-}=\mathcal{H}_{Q}-H^{+}$ where we have used $\mathcal{H}_{Q}$ to be the Hilbert space over the $Q$ physical qubits. The basic premise behind NAED is that a detectable error on any of the physical qubits will move a state $|\psi\rangle$ from $H^{+}$ to $H^{-}$. If the final state is measured and found to be in $H^{-}$, an error is declared. If not, then we measured something in $H^{+}$ and can identify that the logical qubit was measured in the $|0\rangle_L$ or $|1\rangle_L$ state. We stress that this scheme can produce false positives but no false negatives.

In this paper, we will describe a procedure to implement a simple NAED code. This includes the operators required to encode the initial states as well as methods to create logical gates out of the standard base set of $U_3$ and $C_x$ gates. These logical gates are then used to produce logical  Greenberger–Horne–Zeilinger (GHZ) states with differing amounts logical qubits and physical qubits per logical qubit. In general, constructing logical gates remains an active area of research for all error mitigation schemes \cite{Logic1,Logic2,Logic3,Logic4}.

The unencoded GHZ circuit can be thought of as generalized bell circuit and produces the state

\begin{equation}
    GHZ(N,1)=\frac{|00...0\rangle+|11...1\rangle}{\sqrt{2}}
\end{equation}

\noindent Here, there are $N$ $0$s and $N$ $1$s in each ket and the $1$ in $GHZ(N,1)$ represents the fact that there is no encoding with one physical qubit per logical qubit. We then run these circuits and compare the results with and without error detection.

\section{Bit-flip error detection}

The bit-flip encoding (based on classical Hamming Codes \cite{Hamming1}) is the simplest NAED code. In general, a bit-flip encoding on $Q$ physical qubits can be defined by a set $S\subseteq\{0,1,...,Q-1\}$. We then define the two integers

\begin{equation}
x= \left\{
        \begin{array}{ll}
            0 & \quad S=\emptyset \\
            \sum_{i\in S}2^i & \text{otherwise}
        \end{array}
    \right.
\end{equation}

\begin{equation}
y=2^Q-1-x
\end{equation}

\noindent The code words are then defined as $|0\rangle_L=|x\rangle$ and $|1\rangle_L=|y\rangle$ where $x$ and $y$ are written in binary. From this, it is apparent that for $Q$ physical qubits, there are $2^Q$ encodings that we might utilize. For example, a simple set of codewords for $Q=3$ physical qubits is $|0\rangle_L=|001\rangle$ and $|1\rangle_L=|110\rangle$ as determined by the set $S=\{0\}$.

A word on the notation used throughout this paper: let $\sigma_x$ be the standard $2\times 2$ Pauli $x$ matrix. We will use the notation $\sigma_i$ to represent the $2^Q\times 2^Q$ matrix

\begin{equation}
    \sigma_i=I_{2^{i}}\otimes \sigma_x \otimes I_{2^{Q-i-1}}
\end{equation}

\noindent where $I_n$ is the $n\times n$ identity matrix. For example, for $Q=3$ we have

\begin{equation}
    \sigma_0=\sigma_x\otimes I_4
\end{equation}

\begin{equation}
    \sigma_1=I_2\otimes \sigma_x\otimes I_2
\end{equation}

\begin{equation}
    \sigma_2= I_4\otimes \sigma_x
\end{equation}

\noindent Note that qubits are counted starting from $0$ rather than $1$.

For any two physical qubits in the circuit, the gate $C_x(i,j)$ is defined to be a controlled-not gate with control qubit $i$ and target qubit $j$. Additionally, qubits other than $i$ and $j$ are assumed to be operated on by identity gates. The gate $C_x(i,j)$ represents a $2^n\times 2^n$ matrix, rather than a $4\times 4$ matrix, where $n$ will be obvious from context. As an example, for $n=3$ physical qubits

\begin{equation}
    C_x(1,2)=I_2\otimes C_x
\end{equation}

\begin{equation}
    C_x(0,2)=(I_2\otimes SWAP) (C_x\otimes I_2)(I_2\otimes SWAP) 
\end{equation}

\begin{equation}
    C_x(1,0)= (SWAP \cdot C_x\cdot  SWAP)\otimes I_2
\end{equation}

\noindent Here, both $C_x$ and $SWAP$ are the standard $4\times 4$ controlled-not and swap gates respectively. Finally, when using product notation for matrix multiplication, we will use the convention

\begin{equation}
    \prod_{n=1}^N A_n=A_NA_{N-1}\cdots A_{2}A_1
\end{equation}

\subsection{Encoding, Logical Gates, and Detectable Errors}

In order use the bit-flip encoding, two things are required: a way to encode the initial $|0\rangle_L$ state and a method to construct logical gates that operate on $|0\rangle_L$ and $|1\rangle_L$. To encode $|0\rangle_L$ for a given set $S$ (with $Q$ physical qubits starting in the state $|00...0\rangle$) we simply need to operate on each qubit $q_{i}$ with a $\sigma_x$ if $i\in S$. This can be written mathematically as 
\begin{equation}
M_S=\prod_{i\in S}\sigma_{i}\label{MSmatrix}
\end{equation}

\noindent where $M_S$ is identified as the encoding matrix.

The construction of logical gates for the bit-flip encoding is slightly more complicated. By logical gates, we mean that a logical $R$ gate (denoted $L_S(R)$) has the property $L_S(R)|\psi\rangle_L=|\phi\rangle_L$ when $R|\psi\rangle=|\phi\rangle$. Here, the subscript $L$ denotes the logical version of non-encoded qubit states $|\psi\rangle$ and $|\phi\rangle$ and $R$ is an arbitrary physical gate. For $Q$ physical qubits and $S=\emptyset$ we have

\begin{equation}
    L_\emptyset(U_3)=\left[\prod_{i=1}^{Q-1} C_x(0,Q-i)\right] U_3\otimes I_{2^{Q-1}}\left[\prod_{i=1}^{Q-1} C_x(0,i)\right]
\end{equation}

\noindent Note that this logical gate uses $2(Q-1)$ physical $C_x$ gates. For a general set $S$ we have

\begin{equation}
    L_S(U_3) = \left\{
        \begin{array}{ll}
            L_\emptyset(\sigma_xU_3\sigma_x) & \quad 0\in S \\
            L_\emptyset(U_3) & \quad 0\notin S
        \end{array}
    \right.\label{bitflipLU}
\end{equation}

Finally, to define the logical $C_x$ gate we must introduce some notation describing the physical qubits of each logical qubit: let $\{r_0,r_1,...,r_{Q-1}\}$ be the $Q$ physical qubits that correspond to the logical control qubit and let $\{p_0,p_1,...,p_{Q-1}\}$ be the $Q$ physical qubits that correspond to the logical qubit being operated on. Then for an arbitrary $S$ the logical $C_x$ gate is given by

\begin{equation}
    L_S(C_x)=\prod_{i=0}^{Q-1} C_x(r_i,p_i)(I_{2^Q}\otimes M_S)\label{bitflipCx}
\end{equation}

\noindent Here, $M_S$ is the encoding matrix defined in equation \ref{MSmatrix}. Proofs of equations \ref{bitflipLU} and \ref{bitflipCx} are provided in the appendix.

\begin{figure}[H]
    \centering
    \includegraphics[height=3cm]{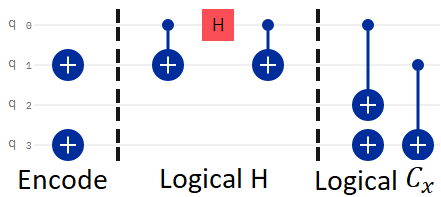}
    \caption{This circuit encodes a bell state for $Q=2$ physical qubits per logical qubit with codewords $|0\rangle_L=|01\rangle$ and $|1\rangle_L=|10\rangle$. The first set of gates to the left of the first barrier transforms $|0000\rangle$ to $|0101\rangle=|00\rangle_L$. The second set of gates is the logical Hadamard gate $L_{\{1\}}(H)$. The third set of gates is the logical $C_x$ gate $L_{\{1\}}(C_X)$. At the end of this circuit, the qubits will be in the linear combination $|\psi\rangle=1/\sqrt{2}(|00\rangle_L+|11\rangle_L)=1/\sqrt{2}(|0101\rangle+|1010\rangle)$.}\label{Bellexample}
\end{figure}

An example of bit-flip encoding and the associated logical gates is given by encoding a bell state with $Q=2$ and $S=\{1\}$. In this example $x=2$ and $y=1$ with $|0\rangle_L=|10\rangle$ and$|1\rangle_L=|01\rangle$. The circuit includes an encoding step, a logical Hadamard $L_{\{1\}}(H)$, and a logical controlled-not $L_{\{1\}}(C_X)$ as shown in FIG \ref{Bellexample}. The final state will be given by $|\psi\rangle=1/\sqrt{2}(|00\rangle_L+|11\rangle_L)=1/\sqrt{2}(|0101\rangle+|1010\rangle)$.

After measurement, there are 16 possibilities: $4$ states in $H^{+}$ and $12$ states in $H^{-}$. If we measure $|0101\rangle$ then we obtain $|00\rangle_L$. In a similar fashion $|1010\rangle$ yields $|11\rangle_L$, $|0110\rangle$ yields $|01\rangle_L$, and $|1001\rangle$ yields $|10\rangle_L$. If we measure any other of the 12 combinations of $0$s and $1$s, then we are in $H^{-}$ and an error is declared.

It is possible to describe the errors that we have a chance to detect while using the bit-flip code. If we model these errors by $2\times 2$ unitary gates, then any error not of the form

\begin{equation}
P(\theta,\phi)=e^{i\phi}\left(
\begin{array}{cc}
 1 & 0 \\
 0 & e^{i\theta} \\
\end{array}
\right)
\end{equation}

\noindent for real $\phi$ and $\theta\neq 2\pi k$, will force any state $|\psi\rangle\in H^{+}$ into some state $|\psi^{'}\rangle\in H^{-}$. These detectable errors include $\sigma_x$, $\sigma_y$, and any other non-phase error that might occur.

\section{An Application of the Bit-Flip Encoding}

The logical $U_3$ and $C_x$ gates are used to construct $GHZ(N,Q)$ circuits where $N$ is the number of logical qubits and $Q$ is the number of physical qubits per logical qubit. Gates corresponding to this circuit are given by

\begin{equation}
    GHZ(N,1)=\left[\prod_{i=0}^{N-1}C_x(i,i+1)\right](H\otimes I_{2^{N-1}})
\end{equation}

\noindent which uses $N-1$ physical $C_x$ gates and a single Hadamard gate. Note that this is not the only way to create the $GHZ(N,1)$ state \cite{betterGHZ}.

\begin{figure}[H]
    \centering
    \includegraphics[height=3cm]{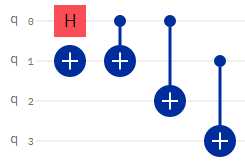}
    \caption{The simplified $GHZ(2,2)$ circuit. This circuit is identical to the circuit in FIG \ref{Bellexample} except that the first $C_x$ gate and bottom $\sigma_x$ gates have been removed as redundant. This does not change the overall state $|\psi\rangle=1/\sqrt{2}(|0101\rangle+|1010\rangle)$ that this circuit produces.}\label{Bellexample2}
\end{figure}

In order to turn the $GHZ(N,1)$ circuit into the $GHZ(N,Q)$ circuit (using bit-flip encoding) we simply replace all physical gates in the $GHZ(N,1)$ circuit with their logical equivalents from equations \ref{bitflipLU} and \ref{bitflipCx}. Of course, we also have to decide which set $S$ we will use for the encoding. In order to balance the number of $0$s and $1$s that make up $|0\rangle_L$ and $|1\rangle_L$, we will use $S=\left\{0,1,...,\lceil Q/2\rceil\right\}$. The last thing we need to do is slightly simplify the resulting circuit before implementation. For example, for the circuit $GHZ(2,2)$ we have the the circuit given in FIG \ref{Bellexample}, however the first $C_x$ gate as well as the two $\sigma_x$ gates on $q_3$ are redundant. Removing these gives us the circuit in FIG \ref{Bellexample2}, and a similar simplification will be used for all experimental runs.

\subsection{Experimental Design, Results, and Discussion}

In order to measure how well our NAED works, we will employ a modified version of the total variation distance \cite{TVD} between probability measure denoted $\tau(A,B)$ for finite probability distribution functions (PDFs) $A$ and $B$. Define the similarity between two finite PDFs to be 

\newpage
$$0\leq \mu(A,B)=100(1-\tau(A,B))$$

\begin{equation}
     =100-50\sum_{i=1}^{|A|}|A_i-B_i|\leq 100\label{similarity}
\end{equation}
     
\noindent The similarity is equal to $0$ if two PDFs are completely dissimilar and $100$ if they are identical. For our experiments we will use the IBMQ quantum computers and compare the experimentally determined PDF for the $GHZ(N,Q)$ circuit with theoretical expected PDF. For the simple $GHZ(N,Q)$ circuit, this theoretical PDF is easily computed and will always be an equally split between two states in the full $\mathcal{H}_{NQ}$ circuit space.

All experiments were performed on the $27$ qubit processor \textit{ibmq\_motreal} \cite{ibmquantum} with a quantum volume \cite{quantumvolume} of $128$. For each $(N,Q)\in \{2,3,4,5\}\times \{1,2,3,4,5\}$, the $GHZ(N,Q)$ circuit was submitted at optimization level $1$ for $2^{13}$ shots. Two similarity measures are computed using equation \ref{similarity}. The first retains all of the experimental results and represents the full similarity measure of the encoded circuit, $\mu_{Full}$. The second omits all experimental results for which an error is detected and is the NAED result, denoted $\mu_{NAED}$. This process was repeated between $220$ and $230$ times for each circuit, and averaged to yield $\mu_{Full}$ and $\mu_{NAED}$ for each $(N,Q)$ pair.

\begin{figure}[H]
    \centering
    \includegraphics[height=6cm]{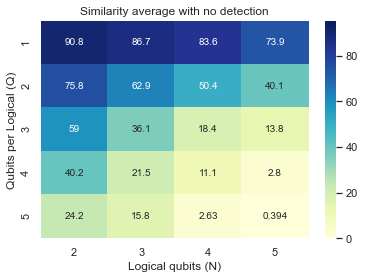}
    \caption{The similarity measure $\mu_{Full}$ of the $GHZ(N,Q)$ circuits over the input space $(N,Q)\in \{2,3,4,5\}\times \{1,2,3,4,5\}$. Not surprisingly, the best results occur at $GHZ(2,1)$ with a similarity measure of $90.8$. The similarity decreases as both $N$ and $Q$ increase, with the worst similarity of $0.4$ for $N=Q=5$.}\label{Simnodet}
\end{figure}

\begin{figure}[H]
    \centering
    \includegraphics[height=6cm]{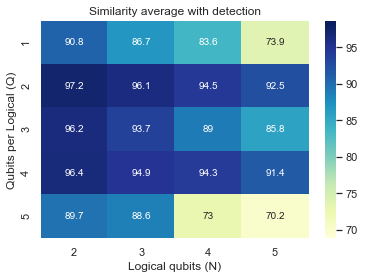}
    \caption{The similarity measure $\mu_{NAED}$ of the $GHZ(N,Q)$ circuits over the input space $(N,Q)\in \{2,3,4,5\}\times \{1,2,3,4,5\}$. The highest similarity is now $97.2$ for $GHZ(2,2)$ with the circuit from FIG \ref{Bellexample2}. while the greatest increase in similarity from the unencoded circuit occurs between $GHZ(5,1)$ and $GHZ(5,2)$.}\label{Simdet}
\end{figure}

Values of $\mu_{Full}$ are shown in FIG \ref{Simnodet} where the best run overall is the $GHZ(2,1)$ circuit with a similarity measure of $90.8$. The similarity then decreases as both $N$ and $Q$ increase. Values of $\mu_{NAED}$ are shown in FIG \ref{Simdet} and

\noindent demonstrate that NAED is a viable option for improving quantum computation. As seen in FIG \ref{Simdet}, values of $\mu_{NAED}$ for $GHZ(N,Q)$ where $2\leq Q\leq 4$ are all greater than $\mu_{NAED}$ for the unencoded $GHZ(N,1)$ circuit. It is not until $Q=5$ that NAED produced lower values of $\mu_{NAED}$ than $Q=1$ where no error detection is performed.

\begin{figure}[H]
    \centering
    \includegraphics[height=6cm]{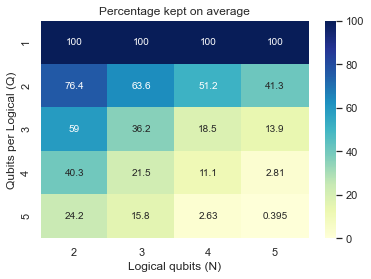}
    \caption{The percentage of runs retained for each $GHZ(N,Q)$ circuit over the input space $(N,Q)\in \{2,3,4,5\}\times \{1,2,3,4,5\}$ after error detection has been performed. For $Q=1$, there are no runs removed. The next highest percentage is for the $GHZ(2,2)$ circuit (given by FIG \ref{Bellexample2}) at $76.4\%$ kept. From here, the percentage retained decreases as both $N$ and $Q$ increase.}\label{Kept}
\end{figure}

Also seen in FIG \ref{Simdet} is evidence that even values of $Q$ outperform odd values of $Q$. For example, the similarity of $GHZ(N,4)$ is greater than the similarity of $GHZ(N,3)$ for all $N$. This behavior is caused by an assymmetry in the number of $0$'s and $1$'s that make up the codewords for odd values of $Q$. Since super conducting qubits will naturally decohere towards the $|0\rangle$ state \cite{deco1,deco2}, this leads to an asymmetry in measured states for odd $Q$.

A primary cost of NAED is the number of runs that are discarded when an error is declared. The total number of runs that are not included in the calculation of $\mu_{NAED}$ increases with $N$ and $Q$ as seen in FIG \ref{Kept} where only a few percent of the runs are retained for larger $N$ and $Q$. It is interesting to note the values of $\mu_{Full}$ in FIG \ref{Simnodet} and percentage retained in FIG \ref{Kept} approach each other in value as both $N$ and $Q$ increase. This observation is used to show that the ratio of false positives to the total number of measurements decreases as $N$ and $Q$ increase. To illustrate why this is the case, define the following

\begin{itemize}
    \item $T = $ the total number of measurements
    \item $r_0=$ the total number of $|00...0\rangle_L$ measured
    \item $r_1=$ the total number of $|11...1\rangle_L$ measured
    \item $r_a = $ the total number of logical states other than $|00...0\rangle_L$ or $|11...1\rangle_L$ measured
    \item $r_b=$ the total number of states which don't fall into any logical state measured

\end{itemize}

\noindent Here, $r_b$ are errors that NAED would catch while $r_a$ are errors that NAED would not catch. For example, with $N=3$ and $Q=2$ the state $|010110\rangle=|001\rangle_L$ is an invalid $GHZ$ state but will not be caught by error detection. Also note that $T=r_0+r_1+r_a+r_b$. Then the percentage of measurements kept is given by

\begin{equation}
P_{Kept}=1-\frac{r_b}{T}
\end{equation}

\noindent Without detection, $\mu_{Full}$ can be written as

\begin{equation}
    \mu_{Full}=1-\frac{1}{2}\left(\left|\frac{1}{2}-\frac{r_0}{T}\right|+\left|\frac{1}{2}-\frac{r_1}{T}\right|+\frac{r_a}{T}+\frac{r_b}{T}\right)\label{eqrr}
\end{equation}

\noindent The absolute values may be ignored since the chance that $\frac{r_0}{T}>\frac{1}{2}$ or $\frac{r_1}{T}>\frac{1}{2}$ is negligible for higher $N$ and $Q$. Then equation \ref{eqrr} becomes

$$
    \mu_{Full}=1-\frac{1}{2T}(T-r_0-r_1+r_a+r_b)
$$

$$
=1-\frac{1}{2T}(T-r_0-r_1-r_a-r_b+2r_a+2r_b)
$$

\begin{equation}
    =1-\frac{r_a}{T}-\frac{r_b}{T}=P_{Kept}-\frac{r_a}{T}\label{xxy}
\end{equation}

As $N$ and $Q$ increase, the value of $r_a/T$ in equation \ref{xxy} goes to zero.

\section{Conclusion}

These experiments have shown that NAED is a viable option for increasing the fidelity of quantum circuits. At its best, it was able to improve the similarity measure of a $GHZ(5,1)$ circuit from $\mu_{Full}73.9$ to $\mu_{NAED}92.5$ using a $GHZ(5,2)$ circuit. Additionally, we have shown that the ratio of false positives to the total number of measurements decreases as the number of logical qubits and the number of physical qubits per logical qubits increase. Current efforts are focused on expanding NAED to other encoding algorithms. As a final comment, NAED may provide higher fidelity quantum communication applications.

\begin{acknowledgments}
The authors would like to thank the Air Force Research Laboratory for providing access to the IBMQ quantum computer as well as Dr. L. D. Merkle for many useful discussions. The views expressed are those of the authors, and do not reflect the official policy or position of IBM or the IBM Quantum team.
\end{acknowledgments}

\appendix*

\section{Proofs of Logical Gates}

\noindent In order to prove the validity of the logical gates described in equations \ref{bitflipLU}-\ref{bitflipCx}, we will use the following four identities:

\begin{equation}
    (I_2\otimes \sigma_x)C_x=C_x(I_2\otimes \sigma_x)\label{relat1}
\end{equation}

\begin{equation}
    (\sigma_x\otimes I_2)C_x(\sigma_x\otimes I_2)=(I_2\otimes\sigma_x)C_x\label{relat2}
\end{equation}

\begin{equation}
    \alpha|0\rangle_L+\beta|1\rangle_L=M_S(\alpha|00...0\rangle+\beta|11...1\rangle)\label{relat4}
\end{equation}

\noindent where $M_S$ is the encoding matrix given in equation \ref{MSmatrix}. The first two of these equations are easily checked while the third equation follows from the definition of $M_S$. For the rest of these proofs, we will drop the $S$ subscript and simply refer to the encoding matrix as $M$. After noting that $M=M^\dag=M^{-1}$, this also gives us

\begin{equation}
    M(\alpha|0\rangle_L+\beta|1\rangle_L)=\alpha|00...0\rangle+\beta|11...1\rangle\label{relat3}
\end{equation}

\subsection{Logical $U_3$ gate}

We will start with the $S=\emptyset$ case (with codewords $|0\rangle_L=|00...0\rangle$ and $|1\rangle_L=|11...1\rangle$) and from there prove the general case. To prove that $L_\emptyset(U_3)$ is a logical $U_3$ gate, we must show that if $U_3(\alpha|0\rangle+\beta|1\rangle)=\tau|0\rangle+\delta|1\rangle$ then $L_\emptyset(U_3)(\alpha|0\rangle_L+\beta|1\rangle_L)=\tau|0\rangle_L+\delta|1\rangle_L$. With the first part of $L_\emptyset(U_3)$ we have

$$
    |\psi_1\rangle=\left[\prod_{i=1}^{Q-1} C_x(0,i)\right](\alpha|0\rangle_L+\beta|1\rangle_L)
$$

\begin{equation}
    =\alpha|00...0\rangle+\beta|10...0\rangle=(\alpha|0\rangle+\beta|1\rangle)\otimes |00...0\rangle
\end{equation}

\noindent where the state $|00...0\rangle$ in the resulting expression has $Q-1$ zeros. Then applying the $U_3$ gate gives us

$$
    |\psi_2\rangle=U_3\otimes I_{2^{q-1}}|\psi_1\rangle=U_3\otimes I_{2^{q-1}}(\alpha|0\rangle+\beta|1\rangle)\otimes |00...0\rangle
    $$
\begin{equation}
    =(\tau|0\rangle+\delta|1\rangle)\otimes |00...0\rangle
\end{equation}

\noindent The final part of $L_\emptyset(U_3)$ gives us

$$
    |\psi_3\rangle=\left[\prod_{i=1}^{Q-1} C_x(0,Q-i)\right]|\psi_2\rangle
$$

$$=\left[\prod_{i=1}^{Q-1} C_x(0,Q-i)\right](\tau|0\rangle+\delta|1\rangle)\otimes |00...0\rangle$$

\begin{equation}
    =\tau|00...0\rangle_L+\delta|11...1\rangle_L=\tau|0\rangle_L+\delta|1\rangle_L
\end{equation}

\noindent as desired.

For the general case, if $0\not\in S$ then

$$    L_S(U_3)(\alpha|0\rangle_L+\beta|1\rangle_L)=L_\emptyset(U_3)(\alpha|0\rangle_L+\beta|1\rangle_L)$$
\begin{equation}    
    =L_\emptyset(U_3)M M(\alpha|0\rangle_L+\beta|1\rangle_L)=L_\emptyset(U_3)M(\alpha|00...0\rangle+\beta|11...1\rangle)\label{stoppingpoint1}
\end{equation}

\noindent We also know that $[L_\emptyset (U_3),M]=\hat{0}$ since every $C_x$ in $L_\emptyset(U_3)$ is controlled by $q_0$, the physical $U_3$ gate in $L_\emptyset (U_3)$ acts on $q_0$, and $0\not\in S$ which allows us to use equation \ref{relat1}. Thus, equation \ref{stoppingpoint1} becomes

$$
    =L_\emptyset(U)M(\alpha|00...0\rangle+\beta|11...1\rangle)$$
    
    $$=M L_\emptyset(U)(\alpha|00...0\rangle+\beta|11...1\rangle)
$$

\begin{equation}
    =M(\tau|00...0\rangle+\delta|11...1\rangle)=\tau|0\rangle_L+\delta|1\rangle_L
\end{equation}

For the case where $0\in S$, let $M^{'}=\sigma_1 M$. Then using the fact $\sigma_i=\sigma_i^{-1}$, we have

\begin{equation}
    L_\emptyset(\sigma_x U_3\sigma_x)M=\sigma_0\sigma_0 L_\emptyset(\sigma_x U_3\sigma_x)\sigma_0 M^{'}
\end{equation}

\noindent Then using equation \ref{relat2} $2(Q-1)$ times (one for each physical $C_x$ in $L_\emptyset(\sigma_x U\sigma_x)$) and simplifying using equation \ref{relat1} we get

\begin{equation}
    =\sigma_0 M^{'}M^{'} L_\emptyset( U_3) M^{'}=ML_\emptyset( U_3)
\end{equation}

\noindent Using this relation we may now conclude

$$    L_S(U_3)(\alpha|0\rangle_L+\beta|1\rangle_L)=L_\emptyset (\sigma_xU\sigma_x)(\alpha|0\rangle_L+\beta|1\rangle_L)
$$

$$
    =L_\emptyset(\sigma_xU\sigma_x)M M(\alpha|0\rangle_L+\beta|1\rangle_L)=M L_\emptyset(U)(\alpha|00...0\rangle+\beta0|11...1\rangle)
$$

\begin{equation}
    =M(\tau|00...0\rangle+\delta|11...1\rangle)=\tau|0\rangle_L+\delta|1\rangle_L
\end{equation}

\subsection{Logical $C_x$ gate} In a similar manner to the previous proof, We will start with the $S=\emptyset$ case and from there prove the general case. To prove that $L_\emptyset(C_x)$ is a logical $C_x$ gate, we must show that $L_\emptyset(C_x)(\alpha|00\rangle_L+\beta|01\rangle_L+\tau|10\rangle_L+\delta|11\rangle_L)=\alpha|00\rangle_L+\beta|01\rangle_L+\delta|10\rangle_L+\tau|11\rangle_L$. As in the main paper, let $\{r_0,r_1,...,r_{Q-1}\}$ be the $Q$ physical qubits the make up the logical control qubit and let $\{p_0,p_1,...,p_{Q-1}\}$ be the $Q$ physical qubits that make up the logical target bit. For ease of notation, the full ket corresponding to these $2Q$ physical qubits shall be written as $|r_0r_1...r_{Q-1};p_0p_1...p_{Q-1}\rangle$, note the semi-colon used to distinguish between both sets of qubits. We then have

$$C_x(r_0,p_0)(\alpha|00\rangle_L+\beta|01\rangle_L+\tau|10\rangle_L+\delta|11\rangle_L)$$

$$
    =C_x(r_0,p_0)(\alpha|00...0;00...0\rangle+\beta|00...0;11...1\rangle
    $$
    $$
    +\tau|11...1;00...0\rangle +\delta|11...1;11...1\rangle)
$$

$$
    =\alpha|00...0;00...0\rangle+\beta|00...0;11...1\rangle
    $$

    \begin{equation}
   +\tau|11...1;10...0\rangle +\delta|11...1;01...1\rangle
    \end{equation}
    
\noindent Repeating this process for the other $Q-1$ physical $C_x$ gates in $L_\emptyset(C_x)$ gives us the desired result. To generalize to all $S$, note that by equations \ref{relat1} and \ref{relat2} we have
    
$$
    L_S(C_x)=L_\emptyset(C_x)(I_{2^Q}\otimes M)=(I_{2^Q}\otimes M)L_\emptyset(C_x)
$$

$$
=( M\otimes I_{2^Q})L_\emptyset(C_x)(M\otimes I_{2^Q})
$$

\begin{equation}
    =( M\otimes M M)L_\emptyset(C_x)(M\otimes I_{2^Q})=( M\otimes M )L_\emptyset(C_x)(M\otimes M)
\end{equation}

\noindent Using this equivalent definition for $L_S(C_x)$, we get

$$
    L_S(C_x)(\alpha|00\rangle_L+\beta|01\rangle_L+\delta|10\rangle_L+\tau|11\rangle_L)
$$

$$
=( M\otimes M )L_\emptyset(C_x)(M\otimes M)(\alpha|00\rangle_L+\beta|01\rangle_L+\delta|10\rangle_L+\tau|11\rangle_L)
$$

$$
    =( M\otimes M )L_\emptyset(C_x)(\alpha|00...000...0\rangle+\beta|00...011...1\rangle)
    $$
\begin{equation}
    +\delta|11...100...0\rangle+\tau|11...111...1\rangle)
\end{equation}

\noindent by equation \ref{relat3}. But this is precisely the relationship we just showed (the $S=\emptyset$ encoding). Thus, it simplifies to

$$
    =( M\otimes M )(\alpha|00...000...0\rangle+\beta|00...011...1\rangle)
    $$
$$
    +\tau|11...100...0\rangle +\delta|11...111...1\rangle)
$$

\begin{equation}
    =\alpha|00\rangle_L+\beta|01\rangle_L+\tau|10\rangle_L+\delta|11\rangle_L
\end{equation}

\noindent as desired.


\bibliography{apssamp}

\end{document}